\documentclass{article}

\usepackage{amsmath}
\usepackage{amssymb}

\title{Alternativity and reciprocity in the
Cayley-Dickson algebra}

\author{S. Kuwata, H. Fujii, and A. Nakashima}

\date{Faculty of Information Sciences, Hiroshima City University, Asaminami-ku,
Hiroshima 731-3194, Japan}

\begin{document}
\maketitle

\newtheorem{notation}{Notation}
\newtheorem{definition}[notation]{Definition}
\newtheorem{lemma}[notation]{Lemma}
\newtheorem{proposition}[notation]{Proposition}

\begin{abstract}
We calculate the eigenvalue $\rho$ of the multiplication mapping $R$ on the
Cayley-Dickson algebra
$\mathbb A_n$. If the element in $\mathbb A_n$ is composed of a pair of alternative
elements in $\mathbb A_{n-1}$, half the eigenvectors of $R$ in
$\mathbb A_n$ are still eigenvectors in the subspace which is isomorphic to $\mathbb
A_{n-1}$.
The invariant under the reciprocal transformation $\mathbb A_n
\times \mathbb A_{n} \ni (x,y) \mapsto (-y,x)$ plays a fundamental role in
simplifying the functional form of $\rho$.
If some physical field can be identified with the
eigenspace of $R$, with an injective map from the field to a scalar
quantity (such as a mass) $m$, then there is a one-to-one map $\pi: m \mapsto \rho$. As
an example, the electro-weak gauge field can be regarded as the eigenspace of $R$,
where
$\pi$ implies that the W-boson mass is less than the Z-boson mass, as in the standard
model. 
\end{abstract}




\section{Introduction}

The Cayley-Dickson algebra $\mathbb A_n$ over a field $F$ is the algebra structure
$F^{2^n}$ given inductively by a pair of $\mathbb A_{n-1}$ through the Cayley-Dickson
process~\cite{schafer}. If the base field $F$ is taken as the real number $\mathbb R$,
$\mathbb A_1, \mathbb A_2$, and
$\mathbb A_3$ correspond to the complex number $\mathbb C$, the quaternion $\mathbb H$,
and the octonion $\mathbb O$, respectively.
For $n < 4$, $\mathbb A_n$ satisfies the alternative law:
$x^2 y = x (xy)$ and $xy^2 = (xy) y$,
so that $\mathbb A_n$ is a composition algebra,
$\| x y \| = \| x \| \| y \|$ for all $x,y$ in $\mathbb A_n$ ($n < 4$), that is, a
normed division algebra.
Thus, the multiplication map $R_x: \mathbb A_n \rightarrow \mathbb A_n$ by $R_x: y
\mapsto yx$ (with $\| x \| = 1$) belongs to $SO(2^n)$ for $n< 4$.

The reciprocal transformation is given by $\mathbb A_n \times \mathbb A_n \ni (x,y)
\mapsto (-y,x)$. Originally, the reciprocal transformation was given by the rotation of
the phase space as $(q,p) \mapsto (-p,q)$, where $q,p$ represent position and
momentum, respectively. As emphasized by Born~\cite{born}, the law of nature is
invariant under the reciprocal transformation.
For example, the bosonic hamiltonian for a harmonic oscillator $\frac{1}{2} (q^2 +
p^2)$, the fermionic one $\frac{i}{2} (qp - pq)$, and the corresponding (anti)commutation
relations ($[q,p] =i, [q,q]=[p,p]=0$) and ($\{ q, p \} = 0, \{ q,q \} = \{ p, p \} =
1$) are all invariant under $(q,p) \mapsto (-p,q)$.
Thus, the invariant under the reciprocal transformation plays a fundamental role in
restricting the theory.

For a physical application, $\mathbb A_n$ has been utilized by many
authors~\cite{gunaydin,soucek,kugo,evans,dixon,lohmus,adler,okubo,gursey}.
However, most of the studies are based on the division algebra (except in
Ref.~\cite{sorgsepp}, where it is discussed that the gravitational force may be embedded
in the enlarged space of $\mathbb A_4$).
One of the reasons of adopting $\mathbb A_n$ for $n < 4$ is that it is closely related
to the (simple) Lie group.
The simple Lie groups are categorized into two types: classical and exceptional ones.
The classical ones, denoted by $SO(n)$, $SU(n)$, and $Sp(n)$, represent the isometric
transformation in a $n$-dimensional vector space over $\mathbb R$, $\mathbb C$, and
$\mathbb H$, respectively.
Note that the isometry over $\mathbb O$ is not a Lie group, due to the lack of
associativity; the multiplication over $\mathbb O$ does not form a group.
However, $\mathbb O$ is still a composition algebra due to alternativity, so that
$\mathbb O$ is related to the simple Lie group of an exceptional type as~\cite{baez}
\begin{align*}
G_2 &\cong {\rm Aut} (\mathbb O), \; F_4 \cong {\rm Isom} (\mathbb O \mathbb P^2), \\
E_6 &\cong {\rm Isom} ((\mathbb C \otimes \mathbb O) \mathbb P^2), \;
E_7 \cong {\rm Isom} ((\mathbb H \otimes \mathbb O) \mathbb P^2), \;
E_8 \cong {\rm Isom} ((\mathbb O \otimes \mathbb O) \mathbb P^2),
\end{align*}
where ${\rm Aut}$, ${\rm Isom}$ and $\mathbb K\mathbb P^2$ represent the
automorphism, isometry and the projective plane over $\mathbb K$, respectively.
For $n > 3$, $\mathbb A_n$ is not directly related to the simple Lie algebra.
This conversely implies that $\mathbb A_n$ (for $n > 3$) is a
natural candidate of an algebra for describing the breaking of the isometry of the
simple Lie group.

The aim of this paper is first to obtain the isometry (or rotational symmetry) breaking
of the multiplication map $R_x$, and then to apply the result to some rotational
symmetry breaking system.
If the field (or state) of the isometry breaking system can be
identified with the eigenspace of $R_x$, with an injective map
from the field and some scalar quantity $q$ (called an ``order parameter''), then it
is found that there is a one-to-one map
$\pi$ from $q$ to the eigenvalue of $R_x$.
As an example, we will show that the electro-weak gauge field can be
regarded as the eigenspace of $R_x$. In this case, the one-to-one map $\pi$ between $q$
(here, the gauge boson mass) and the eigenvalue of $R_x$ implies that (under
some reasonable condition) the W-boson mass is less than the Z-boson mass, as in the
standard model.

In Sec.~2, we first review the basic properties of the Cayley-Dickson algebra, and then
calculate the eigenvalue of $R_{(x,y)}$, to find that half the eigenvectors of
$R_{(x,y)}$ in
$\mathbb A_{n+1}$ (for $n < 4$) are still eigenvectors in the subspace which is
isomorphic to $\mathbb A_n$. This is due to the alternativity of $x$ and $y$ in
$\mathbb A_n$.
To guarantee this property for all $n$, we require that the element in $\mathbb
A_{n+1}$ should be composed of a pair of alternative elements in $\mathbb A_n$.
In Sec.~3, we show how the invariant under the reciprocal transformation simplifies
the functional form of the eigenvalue.
In Sec.~4, we find that the eigenvalue of $R_x$ can be interpreted as the order parameter
in some isometry breaking field, as long as the field can be identified with
the eigenspace of $R_x$.
In Sec.~5, we give summary.

\section{Cayley-Dickson algebra}

\subsection{Basic property}

We review the basic properties of the Cayley-Dickson algebra $\mathbb A_n$
over the real numbers $\mathbb R$.
For $x=(x_1, x_2)$ and $y=(y_1,y_2)$ in $\mathbb R^{2^n} = \mathbb R^{2^{n-1}} \times
\mathbb R^{2^{n-1}}$, the multiplication is inductively given by~\cite{schafer,moreno}
\begin{align}
xy = (x_1 y_1 - \bar{y}_2 x_2, \, y_2 x_1 + x_2 \bar{y}_1), \quad \mbox{with } \bar{x}
= (\bar{x}_1, -x_2),
\label{xy}
\end{align}
so that $\overline{xy} = \bar{y} \bar{x}$ for all $x,y$ in $\mathbb A_n$.
The Euclidean norm and the inner product in $\mathbb R^{2^n}$ are given by
\begin{align*}
\| x \|^2 &= x \bar{x} = \bar{x} x, \\
\langle x, y \rangle &= {\textstyle \frac{1}{2} } (x \bar{y} + y \bar{x}) = \Re (x
\bar{y}),
\end{align*}
respectively,
where $\Re (x)$ represents the real part of $x$, that is, $\Re (x) = \frac{1}{2}
(x+\bar{x})$.
For $n=0,1,2$, and $3$, $\mathbb A_n$ represents $\mathbb R, \mathbb C,
\mathbb H$, and $\mathbb O$, respectively.

\begin{notation}
Denote the commutator by $[x, \, y] = xy - yx$, and the associator by $[x, \, y, \, z]
= (xy)z - x(yz)$.
\end{notation}

The basic properties of $\mathbb A_n$ are summarized in Table~1.
\begin{table}[b]
\caption{Basic properties of $\mathbb A_n$.}
\begin{center}
\begin{tabular}{ccc}
\hline 
$n$ & Property & Identity \\ \hline
$0$ & Self-conjugate & $x = \bar{x}$ \\
$0,1$ & Commutative & $[x, \, y] = 0$ \\
$0,1,2$ & Associative & $[x, \, y, \, z] = 0$\\
$0,1,2,3$ & Alternative & $[x, \, x, \, y] = 0$\\
All  & Flexible & $[x, \, y, \, x] = 0$\\
\hline
\end{tabular}
\end{center}
\end{table}
Due to the flexibility of $\mathbb A_n$, we obtain for all $x,y,z$ in $\mathbb A_n$
the following identity~\cite{moreno}
\begin{align}
\langle x, yz \rangle = \langle x \bar{z}, y \rangle = \langle \bar{y} x, z \rangle.
\label{x,yz}
\end{align}
Moreover, the flexibility leads to~\cite{old}
\begin{align}
\bar{x} (xy) = (y\bar{x}) x,
\label{flex}
\end{align}
which is less popular, compared to the flexible law. However, Eq.~(\ref{flex}) is
useful in obtaining Eq.~(\ref{N}).

\begin{notation}
For $x$ in $\mathbb A_n$, the right and left multiplications are denoted by $R_x$ and
$L_x$, respectively, that is, $R_x, L_x: \mathbb A_n \rightarrow \mathbb A_n$ by
\begin{align*}
y R_x = yx, \quad y L_x = xy.
\end{align*}
\end{notation}
Some may feel somewhat uneasy about the notation like $yR_x$, rather than a usual
notation like $R_x (y)$. If we regard $y \in \mathbb A_n = \mathbb R^{2^n}$ as a
row (not column) vector in $\mathbb R^{2^n}$, and $R_x$ as the corresponding $2^{n}
\times 2^{n}$ matrix, $y R_x = yx$ corresponds to $\langle y| R_x = \langle yx|$ in the
Dirac notation.
From the recursion formula of Eq.~(\ref{xy}), $R_x$ and $L_x$ are decomposed into
\begin{align}
R_x = \left( \begin{array}{cc} R_{x_1} & L_{x_2} \\ - L_{\bar{x}_2} & R_{\bar{x}_1}
\end{array} \right), \quad
L_x = \left( \begin{array}{cc} L_{x_1} & \eta L_{x_2} \\ - \eta
R_{x_2} & R_{x_1} \end{array} \right),
\label{Rx}
\end{align}
where $\eta: \mathbb A_n \rightarrow \mathbb A_n$ by $x \eta = \bar{x}$, so that the
corresponding representation matrix is given by ${\rm diag} \, (1,-1,-1, \ldots,-1)$.
From Eq.~(\ref{Rx}), it is found that
\begin{align}
{}^t R_x = R_{\bar{x}}, \quad {}^t L_x = L_{\bar{x}},
\label{tRx}
\end{align}
where the superscript $t$ represents the transposition.

Now we examine whether $R_x$ (with $ \| x \| =
1)$ belongs to $SO (2^{n})$ or not. In this case, it is sufficient to analyze the
algebra structure of
$R_x R_{\bar{x}}$ instead of $R_x$ itself. This is because we have
$\langle y, y \rangle \rightarrow \langle y R_x, y R_x \rangle = \langle y, (y R_x)
R_{\bar{x}} \rangle = \langle y, y (R_x R_{\bar{x}} ) \rangle$ from Eq.~(\ref{x,yz}).
Since $R_x R_{\bar{x}} \; (= R_x {}^t R_x)$ is positive semidefinite in $\mathbb
R^{2^n}$, its eigenvalue turns out to be (non-negative) real, so that the vector space
$\mathbb R^{2^n}$ can be decomposed into the eigenspace of $R_x R_{\bar{x}}$.
If $y$ is the eigenvector of
$R_x R_{\bar{x}}$ (with its eigenvalue
$\nu$), then we obtain
$\| y \|^2 \rightarrow \nu \| y \|^2$, so that the violation of $R_x \in SO(2^n)$ can
be checked through the relation of $\nu \neq 1$. 
In what follows, we will obtain the eigenvalue
(together with the corresponding eigenvector) of $R_x R_{\bar{x}}$.

\begin{notation}
For $x$ in $\mathbb A_n$, $N_x = R_x R_{\bar{x}}$ and $\bar{N}_x = L_x L_{\bar{x}}$.
\end{notation}
We have the following identity for all $x$ in $\mathbb A_n$:
\begin{align}
N_x = N_{\bar{x}} = \bar{N}_x = \bar{N}_{\bar{x}},
\label{N}
\end{align}
where $N_{\bar{x}} = \bar{N}_x$ is derived from Eq.~(\ref{flex}).
Substituting Eq.~(\ref{Rx}) into $N_x$ and using Eq.~(\ref{N}), we obtain~\cite{kuwata}
\begin{align}
N_x = \left( \begin{array}{cc} A_{x_1,x_2} & -B_{x_1,x_2} \\ B_{x_1,x_2} & A_{x_1,x_2}
\end{array} \right),
\label{Nx}
\end{align}
where $A_{x_1,x_2} = N_{x_1} + N_{x_2}$ and $B_{x_1,x_2} = [ R_{x_1}, \, L_{x_2}]$.
Using the determinant identity $\left| 
\begin{array}{cc} A & -B \\ B & A \end{array}
\right| = | A+iB| | A-iB| = |A+iB| |{}^t A - i{}^t B|$, and noticing that
$A_{x_1,x_2}$ is symmetric and that $B_{x_1,x_2}$ is antisymmetric, we find that
\begin{align}
| N_x - \nu I | = | C_{x_1,x_2} - \nu I|^2,
\label{Nx-nu}
\end{align}
where $C_{x_1,x_2}= A_{x_1,x_2} + i B_{x_1,x_2}$, so that the
eigenvalue of
$N_x$ is equal to that of
$C_{x_1,x_2}$ [while the corresponding eigenvector $r$ of $N_x$, which is doubly
degenerate, is given by
$r=(d_1, d_2)$ and $(d_2, -d_1)$, where $d_1$ and $d_2$ represent the real and
imaginary parts of the eigenvector of $C_{x_1,x_2}$, respectively].

Thus, in the following, we concentrate on the calculation of the eigenvalue of
$C_{x_1,x_2}$.
\begin{notation}
For $x,y$ in $\mathbb A_n$,
we set $C'_{x,y} = C_{x,y} - (\| x \|^2 + \| y \|^2) I$,
where $\| x \|^2 + \| y \|^2$
represents the diagonal elements of $C_{x,y}$, which is obtained inductively.
We also denote by $S_n$ the set of eigenvalues of $C'_{x,y}$.
For later convenience, if the element $s_i \in S_n$ has the same functional form as
$s'_i \in S_{n'} \; (n' \neq n)$, we regard them as the same element.
\end{notation}
For $x,y$ in $\mathbb A_n$ with $n < 4$, $C'_{x,y}$
turns out to be $iB_{x,y}$. For $n < 3$, $B_{x,y}$ is vanishing due to the
associativity of $\mathbb A_n$, so that all the elements of $S_n$ is vanishing. For
$n=3$, $S_n$ is calculated in a somewhat brute-force way~\cite{kuwata}. To summarize, we
have
\begin{align*}
S_0 &= \{ 0 \}, \\
S_1 &= \{ 0,0 \}, \\
S_2 &= \{ 0,0,0,0 \}, \\
S_3 &= \{ 0,0,0,0,
\pm 2 | {\bf x} \times {\bf y} |, \pm 2 | {\bf x} \times {\bf y} | \},
\end{align*}
where $| {\bf x} \times {\bf y} |^2 = \| {\bf x} \|^2 \| {\bf y} \|^2 - \langle {\bf x},
{\bf y} \rangle^2$, with the bold face letter ${\bf x}$ representing the imaginary part
of
$x$, that is, ${\bf x} = x- \Re (x)$.
For $x \in \mathbb A_n$ (with $n < 4$), it is found from Eq.~(\ref{Nx-nu}) that all the
eigenvalues of
$N_x$ are given by $\| x \|^2$, due to the vanishing of all the elements of $S_0,S_1$
and $S_2$. Thus, we obtain
$\| x y \|^2 = \langle x, x N_y \rangle =  \langle x, x
\rangle \| y \|^2 = \| x \|^2 \| y \|^2$ for all $x, y$ in $\mathbb A_n$ (with $n <
4$). This indicates that
$\mathbb A_n$ (for
$n< 4$) is the normed division algebra, as is well known.

It should be noted that the sets $S_n$ (for $n < 4$) satisfy the following inclusion
relation:
\begin{align*}
S_0 \subset S_1 \subset S_2 \subset S_3.
\end{align*}
However, $S_{n-1} \subset S_{n}$ does not, in general hold for $n
\geq 4$ in general; there is a counterexample.
The relation of $S_{n-1} \subset S_n$ implies that half of the eigenvectors of
$N_{(x,y)}$ in $\mathbb A_{n+1}$ are still eigenvectors in the subspace 
$V' = \{ (x,y) = (x_1,x_2,y_1,y_2) \in \mathbb A_{n}^2 = \mathbb A_{n-1}^4 | x_2 = y_2 =
0 \}$, which is isomorphic to $\mathbb A_{n}$.
To be more strict, let $(p,q) = (p_1,p_2; q_1,q_2)$ in $\mathbb A_n^2 = \mathbb
A_{n-1}^4$ be the eigenvector of $N_{(x,y)}$ with its eigenvalue $\nu_{x,y}$, that is,
$(p,q) N_{(x,y)} = \nu_{x,y} (p,q)$. In the subspace $V'$, the eigenequation
turns out to be $(p_1,q_1) N_{(x_1,y_1)} = \nu'_{x,y} (p_1,q_1)$, where
$\nu'_{x,y} = \lim_{x_2,y_2 \rightarrow 0} \nu_{x,y}$.
Suppose that
$[\nu_{x,y} - (\| x \|^2 + \| y \|^2)] \in S_n$ belongs also to $S_{n-1}$. Then, we find
that
$\lim_{x_2,y_2 \rightarrow 0}
\nu_{x,y}$ is the same functional form of $\nu_{x,y}$ where $x$ and $y$ are replaced
by $x_1$ and $y_1$, respectively,
because the element in $S_n$ having the same functional form as the element
in $S_{n-1}$ is regarded as the same.
Thus, we obtain $\nu'_{x,y} = \nu_{x_1,y_1}$, which indicates that the eigenvector of
$N_{(x,y)}$ in $\mathbb A_{n+1}$ is still the eigenvector of
$N_{(x_1,y_1)}$ in the subspace $V' \; (\cong \mathbb A_n)$, as long as $S_{n-1}
\subset S_n$.

In the following subsection, we deal with the case where $S_{n-1} \subset S_n$ is
satisfied for all $n$.
This guarantees that half of the eigenvectors of $N_{(x,y)}$ in
$\mathbb A_{n+1}$ remain the eigenvectors in the subspace which is isomorphic to
$\mathbb A_n$.

\subsection{Alternative entry}

Recall that for $n < 4$, where $S_{n-1} \subset S_n$ is satisfied, $\mathbb A_n$ is
alternative; recall also that for $n \geq 4$, where $S_{n-1} \subset S_n$ does not hold
in general, $\mathbb A_n$ is not alternative.
This implies that if the elements $x,y$ are ``alternative'' in $\mathbb
A_n$, the inclusion relation $S_{n-1} \subset S_n$ is satisfied. Fortunately, this
is indeed true, which will be proven in the following.

\begin{definition}
The element $x$ in $\mathbb A_n$ is alternative, if $[a, \, a, \, x] = 0$ for all $x$
in $\mathbb A_n$.
\end{definition}

\begin{lemma}
For $x,y$ in $\mathbb A_n$ (for $n \geq 3$),
\begin{align*}
{\bf x} \shortparallel {\bf y} \Leftrightarrow B_{x,y} = 0,
\end{align*}
where ${\bf x} = x - \Re(x), \,{\bf y} = y - \Re(y)$, and ${\bf x} \shortparallel {\bf
y}$ represents $| {\bf x} \times {\bf y} | = 0$.
\label{l:x||y}
\end{lemma}

\noindent
{\bf Proof.}
$\Rightarrow$) Since $B_{x,y}$ is independent of $\Re (x)$ and $\Re (y)$, there is no
loss of generality that we can set $y=kx$ (or $x=ky$), where $k \in \mathbb R$. Due to
the flexibility of $\mathbb A_n$, we find that $B_{x,x} = 0$ for all $x$ in $\mathbb
A_n$. Thus, we obtain $B_{x,y} = kB_{x,x} \; (\mbox{or } kB_{y,y}) =0$.

\noindent
$\Leftarrow$) Notice that $B_{x,y} = 0 \Leftrightarrow \forall z \in \mathbb A_n, \,
[y, \, z, \, x] = 0$. Then, we show the contraposition such that if ${\bf x}
\nshortparallel {\bf y}$, then $\exists z \in \mathbb A_n, \, [ y, \, z, \, x ] \neq
0$. From ${\bf x} \nshortparallel {\bf y}$, it follows that both ${\bf x}$ and ${\bf
y}$ are nonzero, so that there is no loss of generality that we can choose ${\bf x} =
e_1$ and ${\bf y} = c_1 e_1 + c_2 e_2$, where $c_1,c_2 \in \mathbb R$ with $c_2 \neq
0$ so as to guarantee ${\bf x} \nshortparallel {\bf y}$. Since $\mathbb A_n$ (for $n
\geq 3$) is not an associative algebra, there always exists an element $z \in \mathbb
A_n$ such that $[e_2, \, z, \, e_1] \neq 0$. Thus, we obtain $\exists z \in \mathbb
A_n, \, [y, \, z, \, x] \neq 0$, where use has been made of $[e_1, \, z, \, e_1] = 0$
by the flexibility of $\mathbb A_n$.
\hspace*{\fill} $\Box$

From Lemma~\ref{l:x||y}, together with Eq.~(\ref{xy}), we find that for $x=(x_1,x_2)$
in $\mathbb A_n = \mathbb A_{n-1} \times \mathbb A_{n-1}$,
\begin{align}
x \mbox{ is alternative} \Leftrightarrow x_1, x_2 \mbox{ are alternative with }
B_{x_1,x_2} = 0.
\label{xis}
\end{align}
The repeated application of Eqs.~(\ref{Nx}) and (\ref{xis}) leads to $N_x = \| x \|^2$
for $x \in \mathbb A_n$ alternative, so that we obtain
\begin{align}
x,y \mbox{ are alternative in } \mathbb A_n \Rightarrow C'_{x,y} = i B_{x,y}.
\label{x,y}
\end{align}

\begin{definition}
For $x,y$ in $\mathbb A_n$, $V(x,y)$ denotes the vector space generated by the set $\{
e_0, x, y, xy \}$, where $e_0$ represents the real element of $\mathbb A_n$. $V^\perp
(x,y)$ denotes the orthogonal complement in $\mathbb A_n = \mathbb R^{2^n}$.
\end{definition}

\begin{lemma}
For $x,y$ are alternative in $\mathbb A_n$, we have
\begin{enumerate}
\item
For ${\bf x} \nshortparallel {\bf y}$, $\dim_{\mathbb R} V(x,y) = 4$;

\item
For all $v \in V(x,y)$, $v B_{x,y} = 0$; 

\item
For all $w \in V^\perp (x,y)$, $w B_{x,y} = - w \tilde{B}_{x,y}$, where
$\tilde{B}_{x,y} := [ R_x, \, R_y ]$.
\end{enumerate}
\label{l:dimV}
\end{lemma}

\noindent
{\bf Proof.}
\begin{enumerate}
\item
The linear independence of $\{e_0,x,y,xy \}$ is equivalent to that of $\{e_0, {\bf x},
{\bf y}, {\bf xy} \}$, so that we can take $x$ and $y$ to be pure imaginary. Suppose
that $c_0 e_0 + c_1 x + c_2 y + c_3 xy = 0$, where $c_i \in \mathbb R$. Then 
from Eq.~(\ref{x,yz}), we obtain
$0=c_0 \langle e_0, e_0 \rangle = c_1 \| x \|^2 + c_2 \langle x,y
\rangle = c_1 \langle x,y \rangle + c_2 \| y \|^2 = c_0 \langle x, y \rangle + c_3 \| x
\|^2 \| y \|^2$. Thus for ${\bf x} \nshortparallel {\bf y}$, we get $c_i=0$
($i=0,1,2,3$), so that the elements in $\{ e_0,x,y,xy \}$ are linearly independent over
$\mathbb R$.

\item
It is sufficient to show that $v B_{x,y}= 0$ for $v=e_0,x,y,xy$. For $v=e_0$, this is
trivial. For $v=x,y$, $v B_{x,y} = 0$ by definition of $x,y$ being alternative.
$(xy) B_{x,y} = - [ y, \, xy, \, x] = [x, \, xy, \, y] = (x(xy))y - x ((xy)y) = (x^2 y)
y - x(xy^2) = x^2 y^2 - x^2 y^2 = 0$, where in the second equality, we have used the
identity $[x, \, y, \, z] = -[z, \, y, \, x]$ for all $x,y$, and $z$ in $\mathbb A_n$
[this identity can be derived from the linearization of the flexibility $[x, \, y, \,
x] = 0$ under $x \rightarrow x + \lambda z$ ($\lambda \in \mathbb R$)].

\item
Since $B_{x,y}$ and $\tilde{B}_{x,y}$ are independent of $\Re (x), \Re(y)$, we can set
$x,y$ to be pure imaginary (notice that $w$ is pure imaginary due to $w \perp e_0$).
From $xw+wx=-(x\bar{w} + w \bar{x}) = - 2 \langle x, w \rangle = 0$, $yw+wy=0$, and
$x(wy)+ (wy)x = x (\overline{yw}) + (wy) x = \bar{x} (\overline{wy}) + (wy)x = 2
\langle \bar{x}, wy \rangle = 2 \langle \bar{x} \bar{y}, w \rangle = 2 \langle xy, w
\rangle = 0$, it follows that $w(B_{x,y} + \tilde{B}_{x,y}) = [x, \, w, \, y] + ((wx)y -
(wy)x) = (xw+wx)y - (x(wy) + (wy)x) = 0$.
\end{enumerate}
\hspace*{\fill} $\Box$

\begin{notation}
For ${\bf x}$ and ${\bf y}$ in $\mathbb A_n \setminus \{ e_0 \}$, $\Delta_{{\bf x},
{\bf y}} := \langle {\bf x}, {\bf y} \rangle I + \frac{1}{2} (R_{\bf x} R_{\bf y} +
R_{\bf y} R_{\bf x} )$.
\end{notation}

In a similar way, as in the proof of Lemma~\ref{l:dimV}, we find that

\begin{lemma}
For ${\bf x}$ and ${\bf y}$ are alternative in $\mathbb A_n \setminus \{ e_0 \}$,
\begin{enumerate}
\item
$\Delta_{{\bf x}, {\bf x}} = \Delta_{{\bf y}, {\bf y}} = 0$;

\item
$\forall v \in V(x,y), \, v \Delta_{{\bf x}, {\bf y}} = 0$.
\end{enumerate}
\label{l:Delta}
\end{lemma}

\begin{notation}
For $x=(x_1,x_2)$ in $\mathbb A_n = \mathbb A_{n-1} \times \mathbb A_{n-1}$, $x_{i_1
i_2 \ldots i_j}$ (for $j=1,2,\ldots$) is given inductively by
$x_{i_1 \ldots i_j} = (x_{i_1 \ldots i_j 1}, x_{i_1 \ldots i_j 2})$ in $\mathbb A_{n-j}
= \mathbb A_{n-j-1} \times \mathbb A_{n-1-j}$. We denote $x_{\underbrace{\scriptstyle 11
\ldots 1}_{k \; \rm times} }$ by $x_{(k)}$.
\end{notation}

For $x,y$ are alternative in $\mathbb A_n$ (with ${\bf x} \nshortparallel {\bf y}$), it
is convenient to decompose the vector space $\mathbb A_n = \mathbb R^{2^n}$ by using
$V(x,y)$ and $V^\perp (x,y)$ as
\begin{align}
\mathbb A_n = \bigoplus_{k=1}^{n-1} V_k,
\label{An}
\end{align}
where
\begin{align*}
V_1 &= \{ r \in \mathbb A_n | r \in V(x,y) \}, \\
V_2 &= \{ r \in \mathbb A_n | r \in V^\perp (x,y); r_1,r_2 \in V(x_1,y_1) \}, \\
V_3 &= \{ r \in \mathbb A_n | r \in V^\perp (x,y); r_1,r_2 \in V^\perp (x_1,y_1);
r_{11},r_{12},r_{21},r_{22} \in V(x_{11},y_{11} \}, \\
\quad \;\; \vdots  \\
V_{n-1} &= \bigcap_{k=0}^{n-3} \{ r \in \mathbb A_n |
r_{i_1 \ldots i_k}  \in V^\perp (x_{(k)}, y_{(k)} ) \} \cap 
\{ r \in \mathbb A_n | r_{i_1 \ldots i_{n-2}} \in V ( x_{(n-2)}, y_{(n-2)} ) \}.
\end{align*}

Now we evaluate $S_n$ for $x$ and $y$ alternative in $\mathbb A_n$. In this case,
$C'_{x,y} = iB_{x,y}$ by Eq.~(\ref{x,y}). Recall that $B_{x,y}$ is anti-symmetric, so
that the eigenpolynomial for $iB_{x,y}$ is an even function, and that the eigenvalue of
$i B_{x,y}$ is a real number.
Thus,
$S_n$ can be decomposed into two sets:
$S_n^+$ and $S_n^-$, where $S_n^+$ ($S_n^-$) represents the non-negative (non-positive)
elements of $S_n$ such that $S_n = S_n^+ \cup S_n^- = S_n^+ \cup (-S_n^+)$.

\begin{notation}
For $x,y$ in $\mathbb A_n$, the set of the eigenvalues of $(iB_{x,y})^2$ is denoted by
$\tilde{S}_n$, so that $S_n^+ \cup S_n^+$ is given by square-rooting the elements of
$\tilde{S}_n$.
\end{notation}
From Lemma~\ref{l:dimV}, it is found that $v \in V(x,y)$ is the eigenvector of $(i
B_{x,y})^2$ with its eigenvalue $0$. Thus, the rest we have to do is to obtain the
eigenvalue of
$(iB_{x,y})^2$ whose eigenvector belongs to $V^\perp (x,y)$. In this case, $(i
B_{x,y})^2$ can be replaced by $(i \tilde{B}_{x,y})^2$, which can be written using
$\Delta_{{\bf x}, {\bf y}}$ as
\begin{align}
(i \tilde{B}_{x,y} )^2 &= (i \tilde{B}_{{\bf x}, {\bf y}})^2 \nonumber \\
&=2 (R_{\bf x} R_{\bf y} R_{\bf y} R_{\bf x} + R_{\bf y} R_{\bf x} R_{\bf x} R_{\bf y}) -
(R_{\bf x} R_{\bf y} + R_{\bf y} R_{\bf x})^2 \nonumber \\
&=4 \left[ | {\bf x} |^2 | {\bf y} |^2 - ( \langle {\bf x}, {\bf y} \rangle -
\Delta_{{\bf x}, {\bf y}} )^2 \right],
\label{iB}
\end{align}
where we have used the $\Re (x)$ and $\Re (y)$-independence of $B_{x,y}$ and
Lemma~\ref{l:Delta}\,i).

The next step is to obtain the recursion formula for $\Delta_{{\bf x}, {\bf y}}$. The
repeated application of Eq.~(\ref{Nx}) yields for $x \in \mathbb A_n$
\begin{align*}
N_x = \| x \|^2 I + \sum_{k=1}^{n-3} \underbrace{\sigma_0 \otimes \sigma_0 \otimes
\ldots  \sigma_0 \otimes}_{k-1 \; \rm times} \, (-i \sigma_2) \otimes \sum_{i_1,
\ldots, i_{k-1}} B_{x_{i_1 \ldots i_{k-1} 1},x_{i_1 \ldots i_{k-1} 2}},
\end{align*}
where $\sigma_0 = \left(
\renewcommand{\arraystretch}{0.5} 
\begin{array}{cc} 1 & 0 \\ 0 & 1 \end{array} \right)$ and
$i \sigma_2 = \left(
\renewcommand{\arraystretch}{0.5} 
\begin{array}{cc} 0 & 1 \\ -1 & 0 \end{array} \right)$.
The linearization of $N_{{\bf x}, {\bf x}} \; (= R_{{\bf x}, \bar{\bf x}} =
-R_{{\bf x}, {\bf x}})$ by ${\bf x} \rightarrow {\bf x} + \lambda {\bf y} \in \mathbb
A_n \setminus \{ e_0 \}$ leads to
\begin{align}
2 \Delta_{{\bf x}, {\bf y}} =
\sum_{k=1}^{n-3} \underbrace{\sigma_0 \otimes \sigma_0 \otimes
\ldots \sigma_0 \otimes}_{k-1 \; \rm times} \, (-i \sigma_2) \otimes \sum_{i_1,
\ldots, i_{k-1} } ( B_{{\bf x}_{i_1 \ldots i_{k-1} 1}, {\bf y}_{i_1 \ldots i_{k-1} 2}} +
( {\bf x}
\leftrightarrow {\bf y}) ).
\label{2Delta}
\end{align}
Recall that we are dealing with the case of $x,y$ being alternative in $\mathbb A_n$.
Then from Lemma~\ref{l:x||y} and Eq.~(\ref{xis}), we find that
${\bf x}_{i_1 \ldots i_{k-1} 1} \shortparallel
{\bf x}_{i_1 \ldots i_{k-1} 2}$, and the analogous relation for ${\bf y}$, so that we can
set
\begin{align}
{\bf x}_{\underbrace{\scriptstyle 11\ldots 1}_{k-1} 2} = \kappa_1^{(k)}
{\bf x}_{(k)}, \quad
{\bf y}_{\underbrace{\scriptstyle 11\ldots 1}_{k-1} 2} = \kappa_2^{(k)}
{\bf y}_{(k)},
\label{x111}
\end{align}
where $\kappa_1^{(k)}, \kappa_2^{(k)} \in \mathbb R$ (for $k=1,2,\ldots, n-3$) are
certain parameters.
Substituting Eq.~(\ref{x111}) into Eq.~(\ref{2Delta}), we obtain
\begin{align}
2 \Delta_{{\bf x}, {\bf y}} = \sum_{k=1}^{n-3} a_{(k)} \prod_{i=1}^{k-1} b_{(i)}
\underbrace{\sigma_0 \otimes \ldots  \sigma_0 \otimes}_{k-1 \; \rm times} \, (i
\sigma_2) \otimes B_{{\bf x}_{(k)}, {\bf y}_{(k)}},
\label{2Delta_xy}
\end{align}
where $\prod_{i=1}^0 b_{(i)}$ stands for $1$, and $a_{(k)}, b_{(i)}$ are given by
\begin{align*}
a_{(k)} = \kappa_2^{(k)} - \kappa_1^{(k)}, \quad b_{(i)} = 1 + \kappa_1^{(i)}
\kappa_2^{(i)}.
\end{align*}
From Eq.~(\ref{2Delta_xy}), the recursion formula for $\Delta_{{\bf x}, {\bf y}}$ can be
written as
\begin{align}
2 \Delta_{{\bf x}_{(k-1)}, {\bf y}_{(k-1)}} = a_{(k)} i \sigma_2 \otimes B_{{\bf
x}_{(k)}, {\bf y}_{(k)}} + 2 b_{(k)} \sigma_0 \otimes \Delta_{{\bf x}_{(k)},
{\bf y}_{(k)}}
\;\; (k=1,2,\ldots,n-3),
\label{recursion}
\end{align}
where ${\bf x}_{(0)}$ denotes $\bf x$ in $\mathbb A_n \setminus \{ e_0 \}$.
Note that Eq.~(\ref{recursion}) is analogous to Eq.~(\ref{Nx}). Recall the
determinant identity $| \sigma_0 \otimes A + i \sigma_2 \otimes B| = | A + iB | | A - iB
|$. Then, we find that all the eigenvalues of $\Delta_{{\bf x}_{(k-1)}, {\bf y}_{(k-1)}
}$ remain invariant, even if $\Delta_{{\bf x}_{(k-1)}, {\bf y}_{(k-1)} }$ is replaced by
\begin{align*}
\Delta_{{\bf x}_{(k-1)}, {\bf y}_{(k-1)}} \rightarrow
\tilde{\Delta}_{{\bf x}_{(k)}, {\bf y}_{(k)}} := \pm {\textstyle \frac{1}{2}}
a_{(k)} i B_{{\bf x}_{(k)}, {\bf y}_{(k)}} + b_{(k)}
\Delta_{{\bf x}_{(k)}, {\bf y}_{(k)}}.
\end{align*}
From Lemmas~\ref{l:dimV} and \ref{l:Delta}, it follows that
\begin{align}
r \tilde{\Delta}_{{\bf x}_{(k)}, {\bf y}_{(k)}} =
\left \{
\begin{array}{ll}
0 & \left[ \,\mbox{for } r \in V(x_{(k)}, y_{(k)}) \right], \\
\pm \frac{1}{2} a_{(k)} \sqrt{(i \tilde{B}_{{\bf x}_{(k)}, {\bf y}_{(k)}} )^2} + b_{(k)}
\Delta_{{\bf x}_{(k)}, {\bf y}_{(k)}} &
\left[ \,\mbox{for } r \in V^\perp(x_{(k)}, y_{(k)}) \right],
\end{array} 
\right .
\label{final}
\end{align}
where $(i \tilde{B}_{{\bf x}_{(k)}, {\bf y}_{(k)}})^2$ can be rewritten using
$\Delta_{{\bf x}_{(k)}, {\bf y}_{(k)}}$ as in Eq.~(\ref{iB}).
Eventually, all the eigenvalues of $\Delta_{{\bf x}_{(k-1)}, {\bf y}_{(k-1)}}$ are given
inductively by those of $\Delta_{ {\bf x}_{(k)}, {\bf y}_{(k)} }$.

\begin{proposition}
For $x,y$ alternative in $\mathbb A_n$, the sets $S_n$ (for $n=0,1,\ldots$) satisfy
the inclusion relation
\begin{align*}
S_0 \subset S_1 \subset S_2 \subset \ldots
\end{align*}
\label{p:inclusion}
\end{proposition}

\noindent
{\bf Proof.}
It is sufficient to show that $\tilde{S}_{n-1} \subset \tilde{S}_n$ for $n=1,2,\ldots$,
because $S_n$ is given by square-rooting the element of $\tilde{S}_n$. In the case
of ${\bf x} \shortparallel {\bf y}$ for $x,y$ in $\mathbb A_n$, where $B_{{\bf x}, {\bf
y}} = 0$ by Lemma~\ref{l:x||y}, all the elements of $\tilde{S}_n$ (hence those of $S_n$)
turn out to be vanishing, so that the Proposition is trivial in this case. In what
follows, we assume that ${\bf x} \nshortparallel {\bf y}$.

We decompose the vector space $\mathbb A_n$ as in Eq.~(\ref{An}).
For all $r \in V_1 = V(x,y)$, where $\dim V_1 = 4$, we obtain $r B_{x,y} = 0$ by
Lemma~\ref{l:dimV}\,ii), so that $V_1$ represents the eigenspace of $(i B_{x,y})^2$ whose
eigenvalue is zero. It should be recalled here that $B_{x,y} = 0$ for all $x,y$ in
$\mathbb A_2$ due to the associativity of $\mathbb A_2$, so that
$\tilde{S}_2 = \{ 0,0,0,0 \}$. Hence, the eigenvalue of $(i B_{x,y})^2$ whose
eigenvector belongs to $V_1$ is given by $\tilde{S}_2$.

If the eigenvector $r=(r_1,r_2)$ in $\mathbb A_n = \mathbb A_{n-1} \times \mathbb
A_{n-1}$ of $(i B_{x,y})^2$ belongs to $V^\perp (x,y)$, its eigenvalue turns out to be
the eigenvalue of
$(i \tilde{B}_{x,y})^2$ by Lemma~\ref{l:dimV}\,iii), so that we have only to calculate
the eigenvalue of $\Delta_{{\bf x}, {\bf y}}$, hence that of $\tilde{\Delta}_{{\bf
x}_1, {\bf y}_1 }$. If $r$ further belongs to $V_2$ (where $\dim V_2 = 4$) so that
$r_1,r_2
\in V(x_1,y_1)$, the eigenvalue of
$\tilde{\Delta}_{{\bf x}_1, {\bf y}_1 }$ turns out to be vanishing by Eq.~(\ref{final})
with $k=1$.
On the other hand, the relation of $\Delta_{{\bf x}, {\bf y}} = 0$, in itself, holds for
all ${\bf x}, {\bf y}$ in $\mathbb A_3$ (where $\dim \mathbb A_3 = 8$) by
Eq.~(\ref{2Delta_xy}) or (\ref{recursion}). Thus, the four eigenvalues of $(i
B_{x,y})^2$ whose corresponding eigenvectors belong to $V_2$ are given by the four
elements in
$\tilde{S}_3$. Recall that among the elements of
$\tilde{S}_3$, the eigenvalue whose eigenvector belongs to $V_1$ is given by
$\tilde{S}_2$, so that the eigenvalue of $(i B_{x,y})^2$ whose eigenvalue belongs
to $V_2$ is given by
$\tilde{S}_3 \setminus \tilde{S}_2$.

In a similar way, we repeatedly apply
Eqs.~(\ref{recursion}) and (\ref{final}) and use the identity $\Delta_{{\bf x}_{(n-3)},
{\bf y}_{(n-3)} } = 0$, to find that the eigenvalue of
$(i B_{x,y})^2$ whose eigenvector belongs to
$V_k$ ($k=2,3,\ldots,n-1$) is given by $\tilde{S}_{k+1} \setminus \tilde{S}_k$, as is
summarized in Table~2.
\begin{table}[b]
\caption{Eigenvalues of $(i B_{x,y})^2$ for $x,y$ alternative in $\mathbb A_n =
\bigoplus_{k=1}^{n-1} V_k$, with ${\bf x} \nshortparallel {\bf y}$.}
\begin{center}
\begin{tabular}{ccc}
\hline
Eigenspace & Eigenvalues & $\dim V_i$ \\ \hline
$V_1$ & $\tilde{S}_2$ & $4$ \\
$V_2$ & $\tilde{S}_3 \setminus \tilde{S}_2$ & $4$ \\
$V_3$ & $\tilde{S}_4 \setminus \tilde{S}_3$ & $8$\\
$\vdots$ & $\vdots$ & $\vdots$\\
$V_{n-1}$  & $\tilde{S}_n \setminus \tilde{S}_{n-1}$ & $2^{n-1}$\\
\hline
\end{tabular}
\end{center}
\end{table}
Thus, we obtain the inclusion relation $\tilde{S}_k \subset
\tilde{S}_{k+1}$ for
$k=2,3,\ldots,n-1$. Since $n$ can be chosen as any $n \geq 3$, we get $\tilde{S}_k
\subset \tilde{S}_{k+1}$ for $k \geq 2$ (for $k=0,1$, the above inclusion relation
is found to be satisfied from the direct calculation of $S_k$).
\hspace*{\fill} $\Box$

In the final analysis, we have found from Eqs.~(\ref{Nx-nu}) and (\ref{x,y}) that for
$x,y$ alternative in
$\mathbb A_n$, the eigenequation for $N_{(x,y)}$ is replaced by
\begin{align*}
r N_{(x,y)} = ( \| x \|^2 + \| y \|^2 + \beta ) r \;\Longleftrightarrow \; d (i B_{x,y})
=
\beta d,
\end{align*}
where $\beta \in S_n$ and $d=r_1 + i r_2$ or $d=r_2 - ir_1$ with $r=(r_1,r_2)$ in
$\mathbb A_{n+1} = \mathbb A_n \times \mathbb A_n$. Noticing that $d (i B_{x,y})^2 =
\beta^2 d$ so that $r_i (i B_{x,y})^2 = \beta^2 r_i$ (for $i=1,2$), and that
$\beta^2 \in \tilde{S}_n \Leftrightarrow \beta \in S_n$,  we find from Table~2 that
\begin{align*}
r_1, r_2 \in V_1 \;& \Longrightarrow \; \beta \in S_2, \\
r_1, r_2 \in V_{k} \; & \Longrightarrow \;\beta \in S_{k+1} \setminus S_{k} \quad
(\mbox{for } k=2,3,\ldots,n-1).
\end{align*}

\section{Reciprocity}

In this section, we show how the invariant under the reciprocal transformation
simplifies the functional form of $\beta \in S_n$. Recall that
$\beta$ is a parameter such that represents the violation of the rotational symmetry
breaking of $R_{(x,y)}$. Actually, we have
$\| r R_{(x,y)} \|^2 = (1+\beta) \| r \|^2$ for $\| x \|^2 + \| y \|^2 = 1$, so that
$\beta = 0$ corresponds to the rotational symmetry of $R_{(x,y)}$.

Before proceeding further, we consider the case of maximal violation of the rotational
symmetry of $R_{(x,y)}$. This consideration will give us some hints of how to
apply the reciprocity.
The maximal violation
is realized for
$\| r R_{(x,y)} \| = 0$, that is, for $\| x \|^2 + \| y \|^2 + \beta = 0$.
This occurs for $\beta \in S_n \setminus S_2$, with
$\beta$ written from Eq.~(\ref{iB}) as
$- 2 \sqrt{ \| {\bf x} \|^2 \| {\bf y} \|^2 - ( \langle {\bf x}, {\bf y}
\rangle - \delta)^2}$,
where $\delta$ represents the eigenvalues of $\Delta_{{\bf x},{\bf y}}$
(the case for
$\beta = +2 \sqrt{\ldots}$ is not appropriate).
Thus from
$( \| x \|^2 + \| y \|^2)^2 = \beta^2$,
$x$ and $y$ should satisfy 
\begin{align}
\Re (x) = \Re (y) = \| {\bf x} \| - \| {\bf y} \| = \langle {\bf x}, {\bf y} \rangle -
\delta = 0.
\label{R(x)}
\end{align}
Here, we have a practical question of which of 
the eigenvalues $\delta$ we should choose.
To restrict the choice of $\delta$, it is convenient to
use the invariance under the reciprocal transformation.
Notice that Eq.~(\ref{R(x)}) is derived from the invariance of
\begin{align}
( \Re (x), \Re (y) ), \; \| {\bf x} \| - \| {\bf y} \|, \; \langle {\bf x}, {\bf y}
\rangle - \delta
\label{Rx,Ry}
\end{align}
under the reciprocal transformation $(x,y) \mapsto (-y,x)$, where $\delta$ is
transformed to $(-\delta)$ due to
$\Delta_{{\bf -y}, {\bf x}} = - \Delta_{{\bf x}, {\bf y}}$.
Furthermore, the invariance of $\| {\bf x} \| - \| {\bf y} \|$ under $(x,y)
\mapsto (-y,x)$ implies the invariance of each term, that is, $\| {\bf x} \|$ and $\|
{\bf y} \|$.
However, the invariance of
$\langle {\bf x}, {\bf y} \rangle - \delta$ does not imply the invariance of each term,
that is,
$\langle {\bf x}, {\bf y} \rangle$ and $\delta$, unless $\langle {\bf x}, {\bf y}
\rangle = \delta = 0$.
Hence, the invariance of $\delta$ leads to $\delta = 0$.
This is our choice of $\delta$ for the case of maximal violation, where $\langle {\bf x},
{\bf y} \rangle = 0$ is required.

Even for a general case, we will require $\langle {\bf x}, {\bf y} \rangle = 0$, which is
obtained from the invariance of $\langle {\bf x}, {\bf y} \rangle$ under $(x,y) \mapsto
(-y,x)$. In a similar way, the invariance of $\langle {\bf x}_1, {\bf y}_1 \rangle$
under $(x,y) \mapsto (-y,x)$ implies $\langle {\bf x}_1, {\bf y}_1 \rangle = 0$.
In this case,
$\Re (x_2) \Re (y_2) = 0$ is required from the decomposition
$\langle {\bf x}, {\bf y} \rangle = (1+b_{(1)}) \langle {\bf x}_{1}, {\bf y}_{1}
\rangle + \Re (x_2) \Re (y_2)$.
Furthermore, the invariance of $\Re (x_2) \Re (y_2) = 0$, namely, $\Re (x_2) = 0$ or $\Re
(y_2) = 0$ under
$(x,y) \mapsto (-y,x)$ implies that $\Re (x_2) = \Re (y_2) = 0$.
In this case, we have
$| {\bf x} |^2 | {\bf y} |^2 = (a_{(1)}^2  + b_{(1)}^2 ) | {\bf x}_1 |^2 | {\bf y}_1
|^2$.
In an analogous way, we obtain from the invariance of the reciprocal transformation,
$\langle {\bf x}_{11}, {\bf y}_{11} \rangle = \Re (x_{12}) = \Re (y_{12}) = 0$ and
so forth, so that
\begin{align}
\left \{
\begin{array}{ll}
\langle {\bf x}_{(k)}, {\bf y}_{(k)} \rangle = 0 & (\mbox{for }k=0,1,\ldots,n-3), \\
| {\bf x}_{(k-1)} |^2 | {\bf y}_{(k-1)} |^2 =
( a_{(k)}^2 + b_{(k)}^2 ) | {\bf x}_{(k)} |^2 | {\bf y}_{(k)} |^2 &
(\mbox{for }k=1,2,\ldots,n-3).
\end{array}
\right.
\label{<x,y>}
\end{align}
Under the condition Eq.~(\ref{<x,y>}), it is convenient to define $\sin \theta_{k}$ and
$\cos \theta_k$ by
\begin{align}
\sin \theta_k = \frac{a_{(k)}}{\sqrt{a_{(k)}^2 + b_{(k)}^2}}, \quad
\cos \theta_k = \frac{b_{(k)}}{\sqrt{a_{(k)}^2 + b_{(k)}^2}}.
\label{cos}
\end{align}
Substituting Eq.~(\ref{<x,y>}) into Eq.~(\ref{final}) and using
Eqs.~(\ref{iB}) and (\ref{cos}), we eventually obtain $\tilde{S}_n$ (whose elements
are quadruply degenerated) in a simple form as
\begin{align*}
\tilde{S}'_3 \setminus \tilde{S}'_2 &= \{ {\it \Delta}^2 \}, \\
\tilde{S}'_4 \setminus \tilde{S}'_3 &= \{ {\it \Delta}_+^2, {\it \Delta}_-^2 \}, \\ 
\tilde{S}'_5 \setminus \tilde{S}'_4 &= \{ {\it \Delta}_{++}^2, {\it \Delta}_{+-}^2, {\it
\Delta}_{-+}^2, {\it \Delta}_{--}^2 \}, \\
& \ldots,
\end{align*}
where $\tilde{S}_n = \bigcup_{i=1}^4 \tilde{S}'_n$ and
${\it \Delta}_{\underbrace{\scriptstyle \pm \ldots \pm}_{k \;{\rm times}}} = {\it \Delta}
\cdot
\cos (\pm \theta_1 \pm \theta_2 \pm \ldots \pm \theta_k)$, with
${\it \Delta} = 2 | {\bf x} \times {\bf y} |$.
Note that
$(x,y) \mapsto (-y,x) \Rightarrow (\kappa_1^{(k)}, \kappa_2^{(k)}) \mapsto
(\kappa_2^{(k)}, \kappa_1^{(k)}) \Rightarrow (a_{(k)}, b_{(k)} ) \mapsto (-a_{(k)},
b_{(k)} ) \Rightarrow \theta_k \mapsto - \theta_k$ for $k=1,2,\ldots,n-3$, so that
${\it \Delta}_{\pm \ldots \pm} \mapsto {\it \Delta}_{\mp \ldots \mp} \; (= {\it
\Delta}_{\pm \ldots \pm})$, as is expected from the invariance of $(i B_{x,y})^2$ under
$(x,y) \mapsto (-y,x)$.

\section{Application}

Now we examine whether or not the simplified $\beta$ is available as an order
parameter in some isometry (rotational symmetry) breaking system.
Consider the electro-weak gauge boson system, where the
$U(1) \times SU(2)$ gauge symmetry is (partially) broken.
The reason of adopting this system is that the gauge fields have a hierarchical
structure ${\bf 1} \oplus {\bf 1} \oplus {\bf 2}$, which is analogous to $V_1
\oplus V_2 \oplus V_3$, as
\begin{align}
\mathbb R^{16} &= \mathbb R^4 \oplus \mathbb R^4 \oplus \mathbb R^8 \nonumber \\
&=
A_\mu \oplus Z_\mu \oplus W_\mu^\pm,
\label{R16}
\end{align}
where $A_\mu, Z_\mu$, and $W_\mu^\pm$ represent the photon, Z-boson, and W-boson
fields, respectively (we treat them as classical, not quantized). At low energy, when
the gauge symmetry is broken, the gauge boson turns out to be massive. Denote by $\phi$
the mass matrix of the gauge boson, that is,
$\phi: \mathbb R^{16} \rightarrow \mathbb
R^{16}$ by $(A_\mu, Z_\mu, W_\mu^+, W_\mu^-) \phi = (0,m_{\rm z} Z_\mu, m_{\rm w}
W_\mu^+, m_{\rm w} W_\mu^-)$, where $m_{\rm z}$ and $m_{\rm w^\pm}$
represent the Z and W-boson masses, respectively.
In this case,
the map from the gauge field to the corresponding mass is injective.
On the other hand, recall that $\mathbb R^{16}$ is decomposed into
\begin{align}
\mathbb R^{16} = V_1 \oplus V_2 \oplus V_3, 
\label{R16=}
\end{align}
where the map from $V_i$ to the eigenvalue of $iB_{x,y}$ is injective, as long as $0,
{\it \Delta}$, and ${\it \Delta}_+$ are not degenerate in magnitude.
Comparing Eqs.~(\ref{R16}) and (\ref{R16=}), we may identify the gauge field with
$V_i$, so that we have the following one-to-one correspondence:
\begin{align*}
\begin{array}{ccc}
(A_\mu, Z_\mu, W_\mu^\pm ) &  \longleftrightarrow & (V_1,V_2,V_3) \\
\phi \downarrow & & \downarrow iB_{x,y} \\
(0, m_{\rm z} Z_\mu, m_{\rm w} W^\pm_\mu) & & (0,  {\it \Delta} \cdot V_2, 
{\it \Delta}_\pm \cdot V_3).
\end{array}
\end{align*}
As a consequence, there exists a one-to-one map $\pi: \mathbb R \rightarrow \mathbb R$
such that
\begin{align*}
(0, m_{\rm z}, m_{\rm w} ) \stackrel{\pi}{\longrightarrow} (0, {\it
\Delta},  {\it \Delta}_+ ),
\end{align*}
where it should be noted that ${\it \Delta}_+ = {\it \Delta}_-$.

The rest we will show is that we can set
$\pi (x) = x^a$ ($a > 0$) under some reasonable assumption.
From the boundary condition $\pi (0) = 0$ and the monomorphism of $\pi (x)$, there exist
$x_0, c_0 \in \mathbb R \setminus \{ 0 \}$ such that $\pi (x_0) = c_0$, so that (after
the scaling of $x \rightarrow x/x_0, \pi (x) \rightarrow \pi (x) / c_0$) there is no loss
of generality that we take $\pi (1) = 1$.
Suppose that if $m_{\rm z}$ scales as $m_{\rm z} \rightarrow \lambda m_{\rm z}$
($\lambda \in \mathbb R$), then $m_{\rm w}$ similarly scales as $m_{\rm w} \rightarrow
\lambda m_{\rm w}$.
This condition can be interpreted, based on the standard model, as follows.
While the gauge-boson mass is proportional to the vacuum expectation value $v$ of the
Higgs field, the ratio
$m_{\rm w} / m_{\rm z}$ is independent of $v$ [it depends on the coupling constants
$(g,g')$ between the Higgs and gauge bosons]. If $v$ and $(g,g')$ can be regarded as
independent parameters, $m_{\rm w} / m_{\rm z}$ remains invariant under the scaling $v
\rightarrow \lambda v$. This is the physical meaning of the scaling
$m_{\rm z} \rightarrow \lambda m_{\rm z}$.
Substituting $m_{\rm z} \rightarrow \lambda m_{\rm z}$ and $m_{\rm w} \rightarrow
\lambda m_{\rm w}$ (with $\lambda = m_{\rm z}^{-1}$) into the relation
$\pi (m_{\rm w}) = {\it \Delta}_+ = \pi (m_{\rm z})  \cos \theta_1$ to eliminate $\cos
\theta_1$, we obtain
$\pi (m_{\rm w}) = \pi (m_{\rm z}) \pi (m_{\rm w} / m_{\rm z})$, where use has been made
of $\pi (1) = 1$.
Suppose further that
this relation holds for all $m_{\rm w}, m_{\rm z}$ in $\mathbb R$. Then, we find from the
boundary conditions $\pi (0) = 0$ and $\pi (1) = 1$, that
\begin{align}
\pi (x) = x^a \quad (\mbox{with } a > 0),
\label{pi}
\end{align}
where $a$ is a constant which amounts to $\left. {\rm d} \pi (x) / {\rm d}x
\right|_{x=1}$.

The prediction of Eq.~(\ref{pi}) is that,
as in the standard model, $m_{\rm w}$ is less than $m_{\rm z}$:
\begin{align*}
\cos \theta_1 =
\frac{\pi (m_{\rm w})}{\pi (m_{\rm z})} =
\left( \frac{m_{\rm w}}{m_{\rm z}} \right)^a.
\end{align*}
It is apparent that $\theta_1 = \theta_{\rm w}$ for $a=1$, where $\theta_{\rm w}$
represents the Weinberg angle, although the value of $a$ in itself is not specified at
the present stage.

\section{Summary}

We have calculated the eigenvalue of the right multiplication $R_{(x,y)}$ for $x$
and $y$ alternative in the Cayley-Dickson algebra $\mathbb A_n$, by calculating,
instead, the eigenvalue $\beta$ of $i B_{x,y}$. The eigenvalue
$\beta$ represents the violation of the rotational symmetry of $R_{(x,y)}$.
The requirement of the alternative entries guarantees that half of the eigenvectors of
$N_x$ in
$\mathbb A_n$ are still eigenvectors in the subspace which is isomorphic to $\mathbb
A_{n-1}$.
In calculating $\beta$, the essential point lies in
Lemma~\ref{l:dimV}\,iii). Here, we comment on this property from a Lie algebraic point of
view.
Denote $[L_x, \, L_y]$ by $\tilde{B}'_{x,y}$. Then, it is analogous to show that $w
B_{x,y} = - w
\tilde{B}'_{x,y}$ for $w \in V^\perp (x,y)$, as in Lemma~\ref{l:dimV}\,iii), so that we
obtain
$w B_{x,y} = - w D_{x,y}$, where $D_{x,y} = B_{x,y} + \tilde{B}_{x,y} +
\tilde{B}'_{x,y}$.
It should be noted that $D_{x,y}$ is a derivation of an alternative algebra. The
derivation $D$ of an algebra $\mathfrak A$ is defined by $(xy)D = (xD)y + x (yD)$ for all
$x,y$ in $\mathfrak A$. If $\mathfrak A$ is an alternative algebra, $D$ is given by the
sum $\sum D_{x_i,y_i}$ for $x_i,y_i$ in $\mathfrak A$~\cite{schafer}.
Thus, $D$ forms a Lie algebra, so that the corresponding Lie group turns out to be
isometric.
Hence, we obtain
$\| w \| = \| w e^{\theta D_{x,y}} \| = \| w e^{-\theta B_{x,y}} \|$
for $w \in V^\perp (x,y)$ and $\theta \in \mathbb R$.
Recall that $v = v e^{- \theta B_{x,y}}$ for $v \in V(x,y)$ by Lemma~\ref{l:dimV}\,ii).
Then, we find that the map
$e^{- \theta B_{x,y}}: \mathbb A_n \rightarrow \mathbb A_n$ is isometric in $\mathbb
A_n$.
The subspace of $\mathbb A_{n+1}$ with (without) the rotational symmetry of $R_{(x,y)}$
in $\mathbb A_{n+1}$ is invariant (transformed) under the isometric transformation
$e^{-\theta B_{x,y}}$ in $\mathbb A_n$.

The reciprocity is useful in simplifying the functional form of $\beta$. Assume that
each term in Eq.~(\ref{Rx,Ry}) is invariant under the reciprocal transformation $(x,y)
\mapsto (-y,x)$. Then, it follows that $\langle {\bf x}, {\bf y} \rangle = 0$.
In this case, together with some other invariants,
$\beta$ is simply written as
${\it \Delta}_{\pm \ldots \pm}$.
This simplified expression would reduce further analysis.
In connection to (infinite dimensional) Lie algebra, it may be interesting to point out
that one of the solutions (called $R$ matrix) to the Yang-Baxter equation is written
using
${\it
\Delta}_{\pm
\ldots \pm}$ as
\begin{align*}
R= \frac{1}{\it \Delta}\left( \begin{array}{cccc}
{\it \Delta}_+ & 0 & 0 & 0 \\
0 & {\it \Delta}_{++} & {\it \Delta}_{+++} & 0 \\
0 & {\it \Delta}_{+++} & {\it \Delta}_{++} & 0\\
0 & 0 & 0 & {\it \Delta}_{+}
\end{array} \right),
\end{align*}
with $\theta_1 + 2 \theta_2 + \theta_3 = 0$.

If some physical field $S$ with an isometry breaking has a hierarchical
structure
${\bf 1} \oplus {\bf 1} \oplus {\bf 2} \oplus {\bf 4} \oplus \ldots$, with an injective
map from the field to a scalar quantity $m$ (called an ``order parameter''),
then there is a one-to-one map $\pi: m \mapsto \beta$.
One example is realized in the electro-weak gauge boson field, where
${\bf 1} \oplus {\bf 1} \oplus {\bf 2} = A_\mu \oplus Z_\mu \oplus W_\mu^\pm$, with an
injective map from the gauge field to the corresponding
mass.
In this case,
the one-to-one map $\pi$ implies that (under some reasonable
condition) the W-boson mass is less than the Z-boson mass, as in the standard model.

Another example may be found in the light meson
field, where the $SU(3)$ flavor symmetry is breaking.
The octet ${\bf 8}$ in the light mesons
${\bf 3} \otimes \bar{{\bf 3}} \;( = {\bf 8} \oplus {\bf 1}$), 
forming a 32-dimensional vector space, has a hierarchical structure
\begin{align*}
{\bf 8} &=
{\bf 1} \oplus {\bf 1} \oplus {\bf 2} \oplus {\bf 4} \\
&=
 \eta \oplus \pi^0 \oplus (\pi^+,\pi^-) \oplus (K^+, K^-, K^0, \bar{K}^0 ),
\end{align*}
where $\eta, \pi$, and $K$ represent the corresponding meson fields.
Recalling that $\mathbb R^{32}$ can be decomposed into $V_1 \oplus V_2 \oplus V_3
\oplus V_4$, we find that there is a one-to-one correspondence between $\beta$ and
the meson mass. The detailed analysis of the meson mass is beyond the scope of the
present paper, which, however, will be discussed elsewhere.

\section*{Acknowledgments} The authors are indebted to K.~Terada and M.~Terauchi for
their stimulating discussion.

\end{document}